\documentclass[aps,twocolumn,prl]{revtex4}
\usepackage{amsfonts}
\usepackage[final,hiresbb]{graphicx}
\usepackage{amsmath}
\usepackage{amssymb}
\usepackage{amstext}
\usepackage{epstopdf}
\usepackage{makeidx}

\usepackage{graphicx,color}

\makeindex

\begin{document}
\title{A Single Mobility Function for the Square-Lattice Ising Model and Its Application to Calibrated Monte Carlo Kinetics}
\author{Liangzhe Zhang$^{a}$, Timothy Bartel$^{b}$, Mark T. Lusk$^{a,}$\footnote{Corresponding author. Tel.: 303-273-3675; fax: 303-273-3919. E-mail: mlusk@mines.edu (M.T. Lusk)}}
\affiliation{$^{a}$Department of Physics, Colorado School of Mines, Golden, CO
80401\\
$^{b}$Sandia National Laboratory, Albuquerque, NM 87185}

\begin{abstract}
Computational experiments are used to show that grain boundary mobility is independent of driving force in a two-dimensional, square-lattice Ising model with Metropolis kinetics. This is established over the entire Monte Carlo temperature range. A calibration methodology is then introduced which endows the Monte Carlo algorithm with time and length scales and allows the Monte Carlo parameters to be expressed directly in terms of experimentally measurable parameters. A comparison of results obtained for a variety of driving forces and temperature conditions indicates that such Calibrated Monte Carlo models are able to capture the grain boundary kinetics predicted by sharp-interface theory.

{\bf $Keywords:$} Calibrated Monte Carlo, grain boundary, mobility, driving force, Ising model, anisotropy, surface energy
\end{abstract}
\maketitle

\section{Introduction}
Meso-scale simulation of polycrystalline grain coarsening can be efficiently carried out using Monte Carlo (MC) algorithms. Each site represents many millions of atoms of like orientation, and this orientation is tracked via a discrete, non-conserved order parameter\cite{Holm20012981,Rollett200169}. As opposed to the atomistic kinetic Monte Carlo (KMC) methods \cite{yang.031910}, meso-scale MC models have no intrinsic time and length scales. However, they can be endowed with such physicality by developing a parametric correspondence with sharp-interface (SI) properties which can be experimentally measured. The result is a \emph{Calibrated Monte Carlo (CMC) algorithm}~\cite{PhysRevE.66.061603}. 

%If a grain boundary is idealized as a sharp interface (SI), then its rate of normal accretion can be expressed as the sum of pairs of driving forces multiplied by their respective mobilities. It is typical to consider two such pairs--that associated with interfacial energies and those collectively associated with differences in bulk energy on either side of the grain boundary. The former pair is known as the grain boundary stiffness. Both the driving forces and their mobilities can individually depend on grain boundary inclination--i.e. they can be anisotropic.  The mobilities and driving forces can be measured experimentally, and this provides the basis for the required temporal and spatial calibrations. 

The calibration takes advantage of the fact that MC and SI paradigms both exhibit a linear relation between grain boundary velocity, $\overline{v}$, and total driving force, $\overline{P}$ ~\cite{Viswanathan19731099,Huang19992259}. Within the deterministic SI setting, this is written as
\begin{equation}
\overline{v}=\overline{M}\  \overline{P},
\label{2dvelocity}
\end{equation}%
with $\overline{M}$ the grain boundary mobility. The kinetic relation can be nondimensionalized using  a characteristic length, $l_{0}$,  time, $t_{0}$, and energy, $E_{0}$. The nondimensional kinetic quantities,  
\begin{eqnarray}
v&=&\frac{\overline{v}t_{0}}{l_{0}} \notag \\
 M&=&\frac{\overline{M}E_{0}t_{0}}{l_{0}^{3}}  \notag \\
P &=&\frac{\overline{P}l_{0}^{2}}{E_{0}}, 
\label{nondim}
\end{eqnarray}%
then give
\begin{equation}
v=MP.
\end{equation}%
Here and subsequently, parameters without a bar are understood to be nondimensional. In general, driving forces are generated by either capillary or bulk effects. Curvature forces are described by~\cite{Herring.1949, Lobkovsky2004285}
\begin{equation}
P_{\kappa }=(\gamma +\gamma ^{^{\prime \prime }})\kappa ,  \label{Herring}
\end{equation}
where $\gamma $ is the surface energy, $\gamma ^{^{\prime \prime }}$ is the second derivative of the surface energy  with respect to inclination angle, and $\kappa $ is the grain boundary curvature; $(\gamma +\gamma ^{^{\prime \prime }})$ is the \emph {grain boundary stiffness}~\cite{PhysRevLett.48.368}, which is also written as $\gamma^{\ast}$. The addition of a driving force due to a bulk energy difference per unit volume, $b$, across the boundary gives the general kinetic relation
\begin{equation}
v=M \gamma^{\ast} \kappa +Mb,  \label{bulk}
\end{equation}
This expression is valid provided that grain boundary mobility is independent of the types of driving forces~\cite{Viswanathan19731099,Huang19992259}. The experimentally measurable quantities which must be supplied to this continuum, sharp-interface model are $M$, $b$, and $\gamma$. These quantities can be used to calibrate MC models.

Now consider a MC approach to study the same problem. A two-state field distinguishes grain orientation on a fixed lattice with microstructural evolution described as a Markov chain~\cite{Landau.2005}.  A nondimensional interfacial free energy can be numerically calculated or analytically derived. This free energy along with bulk energy differences can be used to construct MC driving forces. There are many ways in which a spin lattice can be evolved, but the Metropolis algorithm~\cite{Landau.2005} is thought to properly capture the essence of thermally activated events. A lattice site is chosen at random and is temporarily given a new orientation. The probability of accepting such a trial fluctuation of the system is made proportional to an Arrhenius relation involving the energy change and a non-physical Monte Carlo temperature, $T_{mc}$. An MC time step, $\tau_{mc}$, can then be defined as a set of N such trials with N the number of lattice sites. Computational experiments can thus be contrived to measure the proportionality between accretive MC speed and an isolated type of driving force--i.e. to measure MC mobilities. Such mobilities will, of course, be a function of MC temperature and could, in principal, depend on the driving force as well. The MC mobilities may be influenced by the underlying MC lattice and should not be assumed to naturally capture the inclination dependence of the physical systems they mimic.  If the proper inclination dependence is not captured, though, a MC model will not produce morphological evolutions which correspond to experimental measurements. 

A significant number of studies conclude that, in fact, morphological evolution is correctly captured by MC simulations~\cite{Landau.2005,Holm20012981,Rollett200169}. This suggests that the MC mobilities do capture an inclination dependence which matches reality. The implication is that if a calibration can be made for a prescribed boundary geometry then it will apply to arbitrary geometries as well. Consideration must therefore be given to the functional form of the mobilities since the thermodynamic driving forces can be either derived or numerically estimated.

Recent investigations have made important inroads towards understanding the mathematical structure of MC mobilities and have set the stage for more comprehensive numerical work. Lobkovsky et al.~\cite{Lobkovsky2004285,Mendelev.2002,Mendelev20003711} studied grain boundary mobility with a two-dimensional Ising model~\cite{McCoy.1973}  on a triangular lattice. The grain boundary stiffness, which comprises the capillary driving force, was derived under the assumption that the MC temperature is sufficiently low that a kink model of the interface is valid. To understand the methodology, consider a flat boundary with length, $L$, and inclination angle, $\varphi$. The number of geometrically necessary kinks is
\begin{equation}
K=\frac{L}{a\sqrt{3}/2}\sin \varphi ,
\end{equation}
where $a\sqrt{3}/2$ is the triangular kink height. Thermally generated kinks can be neglected at low temperatures, and the total number of lattice sites between the geometrically necessary kinks is then
\begin{equation}
N=\frac{L}{a}\cos \varphi -\frac{1}{2}K.
\end{equation}
The resulting configurational entropy of the flat boundary is
\begin{equation}
S_{\gamma }=\ln \frac{(N+K)!}{N!K!},
\end{equation}
and the grain boundary stiffness, $\sigma^*$, is therefore~\cite{Lobkovsky2004285}
\begin{equation}
\gamma^* = \gamma +\gamma ^{^{\prime \prime }}=-T(S_{\gamma }+S_{\gamma }^{^{\prime
\prime }})=\frac{2\sqrt{3}T}{a\sin 3\varphi }.
\end{equation}
This stiffness does not endow the MC model with a time scale though. That is accomplished in two pieces.  First a grain boundary mobility is approximated using the kink drift velocity~\cite{Lobkovsky2004285} to be
\begin{equation}
M=\frac{\sqrt{3}a^{3}}{2\tau}\frac{\varphi }{T}.
\end{equation}
Here $\tau$ is the nondimensional time associated with one MC step. The remaining piece of the calibration is to prescribe a physical time to each MC step. In practice, this time calibration could be accomplished by comparing the rate of MC boundary evolution with that measured experimentally.  From this point, a relationship can be established between the dimensional sharp-interface parameters and the MC parameters. The resulting CMC code would allow experimental parameters to be used directly in the MC paradigm and would endow the Markov chain with a physical time scale.

Lobkovsky et al. showed that both the mobility and stiffness are anisotropic but that the reduced mobility is isotropic ~\cite{Lobkovsky2004285}. This suggests that there may only be a single MC mobility to contend with and that its dependence on inclination angle can be derived from that of the grain boundary stiffness. This is consistent with an earlier computational study on a 2-D square lattice~\cite{Holm20012981} where the reduced mobility was found to be isotropic as well. In a subsequent numerical study at a single (low) temperature $(T_{mc}/J=0.2)$, Lobkovsky et al. found the mobility to be independent of driving force~\cite{Lobkovsky2004285}. 

In the current work, we build on these studies within the setting of 2-D square lattices. Consistent with Lobkovsky et al. ~\cite{Lobkovsky2004285}, we find that the reduced mobility is isotropic. It is numerically measured under the action of capillary forces, and the mobility is then backed out using an inclination-dependent analytical solution for the grain boundary stiffness. This mobility is then applied to study planar boundaries with a range of inclination angles that are driven only by bulk energy differences. We find that, over a broad range of MC temperatures, the mobility is independent of driving force. This extends the conclusions of Lobkovsky and co-workers to the square lattice system and to the full spectrum of MC temperatures. The results lend additional weight to possible universality of two key properties: that reduced mobilities are isotropic; and, there is only a single mobility that is independent of driving force. The assumption is then made that both conclusions are valid for experimental boundaries in order to focus on calibration methodology. Obtaining the characteristic time, the only non-trivial effort, is accomplished using a computationally measured MC mobility for a single inclination. The work extends our previous results~\cite{PhysRevE.66.061603} by removing the restriction to high MC temperatures where isotropy can be assumed.

%Both the stiffness, $(\gamma
%+\gamma ^{^{\prime \prime }})$, and the mobility, $M$, are anisotropic. However, the product of these quantities, 
%typically referred to as the \emph{reduced mobility}, $M^{\ast }$, is isotropic~\cite%
%{Lobkovsky2004285}:

%%
%\begin{equation}
%M^{\ast }=M(\gamma +\gamma ^{^{\prime \prime }})\doteq \frac{a^{2}}{\tau _{m}%
%}.
%\end{equation}
%

%Lobkovsky et al. also measured grain boundary mobility via MC simulation
%with effects of different types of driving forces, and found it to be independent of the types of driving forces. 
%However, this computation was carried out at a single low temperature. 

%In the current work, we pursue a similar path for the Ising model on square lattice. As in the case of the triangular lattice, an
%analytical solution of surface energy is available. Therefore, an exact expression 
%for the grain boundary stiffness can be obtained as a function of boundary inclination 
%angle and MC temperature. Here, though, the analytical solution is valid for values of MC temperature. The consideration of finite temperature, though, precludes the use of a kink drift velocity to construct the form of the MC mobility function. Instead, computational experiments can be performed to obtain this relation. The purely computational approach, while less elegant, can be immediately extended to three dimensions.

\section{The Square Lattice Ising Model}
The two-dimensional, square lattice Ising system is characterized by an interfacial energy that varies strongly with inclination angle--i.e. is anisotropic~\cite{PhysRev.82.87}. For an interface with zero inclination angle, Onsager derived the interfacial energy as a function of temperature~\cite{PhysRev.65.117}:
\begin{equation}
\frac{\sigma (\alpha)\Delta}{J}= 2-\alpha \ln \{\coth (\frac{1}{\alpha })\},
\label{Onsager}
\end{equation}
where $\alpha =T_{mc}/J$ is the effective temperature, $\Delta$ is the side length of one lattice cell, and $T_{mc}$ and $J$ are the fundamental temperature and interaction energy, respectively. This analytical expression for surface energy was subsequently extended by Abraham to allow for an arbitrary inclination angle, $\varphi$,~\cite{PhysRevLett.33.377,PhysRevB.24.6274, PhysRevLett.15.621,Weeks.1977}:
\begin{equation}
\frac{\sigma(\varphi ,\alpha)\Delta}{J}={\alpha}(\eta _{1}\cos \varphi +\eta _{2}\sin \varphi ),
\label{gamma}
\end{equation}
where
\begin{eqnarray}
\eta _{1} &=&\sinh ^{-1}[a(\varphi ,\alpha)\cos \varphi ]  \notag \\
\eta _{2} &=&\sinh ^{-1}[a(\varphi ,\alpha)\sin \varphi ]  \notag \\
a(\varphi ,\alpha) &=&A[1-(2/A)^{2}]^{1/2} \notag\\
                      && \times\{1+[\sin ^{2}2\varphi +(2/A)^{2}\cos^{2}2\varphi ]^{1/2}\}^{-1/2}  \notag \\
A &=&\cosh ^{2}2K/\sinh 2K  \notag \\
K &=&1/\alpha .
\end{eqnarray}

\begin{figure}[ptb]\begin{center}
\includegraphics[width=0.4\textwidth]{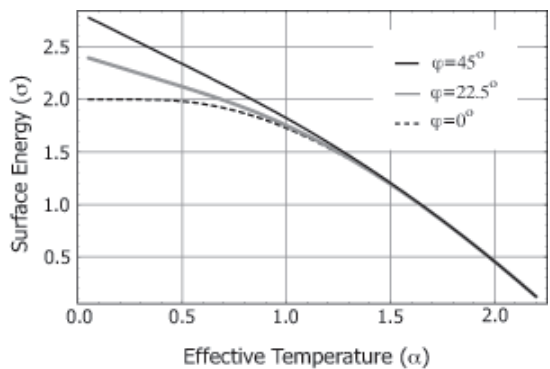}
\caption{
Temperature dependence of surface energy for three inclination angles, $\varphi$.
}
\label{2d_surface_eng}
\end{center}\end{figure}

\begin{figure}[ptb]\begin{center}
\includegraphics[width=0.4\textwidth]{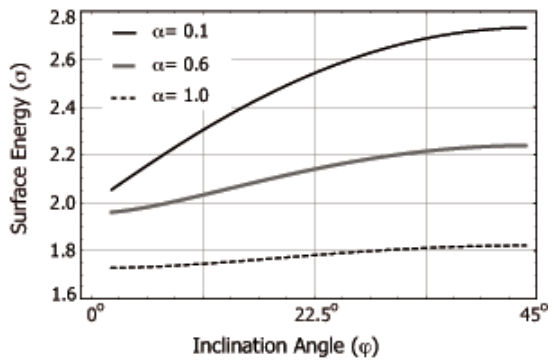}
\caption{
Angle dependence of surface energy at three effective temperatures, $\alpha$.
}
\label{2d_surface_eng_ang}
\end{center}\end{figure}

\begin{figure}[ptb]\begin{center}
\includegraphics[width=0.40\textwidth]{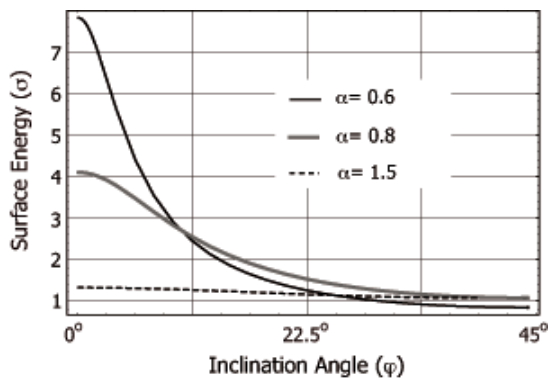}
\caption{
Angle dependence of grain boundary stiffness at three effective temperatures, $\alpha$.
}
\label{2d_stiffness}
\end{center}\end{figure}

\begin{figure}[ptb]\begin{center}
\includegraphics[width=0.4\textwidth]{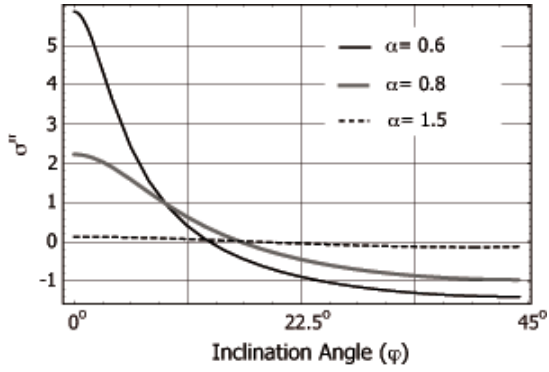}
\caption{
Angle dependence of the second derivative of surface energy with respect to inclination angle. This quantity is plotted for three values of effective temperature, $\alpha$.
}
\label{2d_sureng_scnd}
\end{center}\end{figure}

Figs.~\ref{2d_surface_eng} and \ref{2d_surface_eng_ang} show the dependence of this energy on inclination angle and temperature, respectively. Eq.~(\ref{gamma}) can then be used to derive grain boundary stiffness of the Ising model~\cite{Akutsu.1986}:
\begin{eqnarray}
\sigma^*&=&\frac{(\sigma +\sigma^{^{\prime \prime }})\Delta}{J}=\frac{{\alpha}a(\varphi ,\alpha)}{(\eta _{3} +\eta _{4})}   \notag \\
\eta _{3} &=&\sin ^{2}\varphi(1+\cos^{2}\varphi a^{2})^{1/2}   \notag \\
\eta _{4} &=&\cos ^{2}\varphi (1+\sin ^{2}\varphi a^{2})^{1/2}.
\label{stiffness}
\end{eqnarray}
Figs.~\ref{2d_stiffness} and \ref{2d_sureng_scnd} show the grain boundary stiffness, $\sigma^*$, and the second derivative of surface energy, $\sigma^{^{\prime \prime }}$, as functions of inclination angle for three values of effective temperature, $\alpha$.  At low temperatures, $\sigma^{^{\prime\prime }}$ dominates the stiffness and it strongly depends on inclination angle. At high temperatures, though, the stiffness is almost isotropic since $\sigma^{^{\prime \prime }}$ is negligible. The grain boundary stiffness is then well approximated by the surface energy $\sigma$.
 

With an analytical expression for the driving forces in hand, the next step is to computationally determine the MC mobility function. This is undertaken within a purely capillary setting:
\begin{equation}
v=M\gamma^{\ast} \kappa =M^{\ast }\kappa .  \label{capillary}
\end{equation}
A second configuration can then be used to measure the speed of a planar boundary driven only by a difference in bulk free energy:
\begin{equation}
v=M\ b,  \label{bulkeng}
\end{equation}
The MC mobility can be computed here as well and compared with the capillary result to determine the degree to which the mobility is independent driving force.

\section{Monte Carlo Mobility for the 2-D Square Lattice}
\begin{figure}[ptb]\begin{center}
\includegraphics[width=0.35\textwidth]{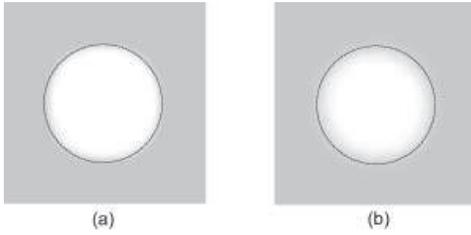}
\caption{
Monte Carlo simulation of inner grain shrinking at (a) low temperature without bulk energy, (b) high temperature without bulk energy. The grid size is $200\times 200$, and 100 experiments were averaged. The dotted circles are a guide to the eye.
}
\label{mc_result}
\end{center}\end{figure}

\begin{figure}[ptb]\begin{center}
\includegraphics[width=0.4\textwidth]{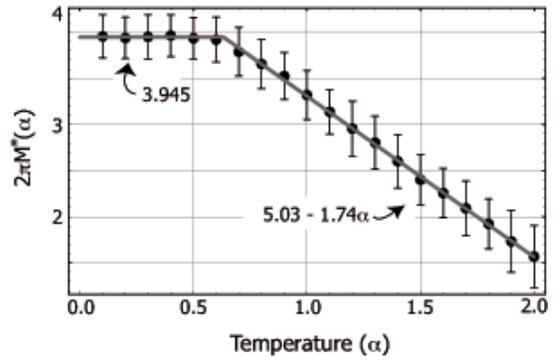}
\caption{
Numerically derived inner grain shrink rate and bi-linear fit. A $200\times 200$ grid was used with an initial internal area fraction of 0.5. Each point represents the average of 200 experiments, and error bars show the standard deviation of the shrink rate for each temperature. 
}
\label{2d_shrink_rate}
\end{center}\end{figure}

Consider a 2D square grain which encloses a circular inner grain. In the absence of external fields, SI kinetics implies that the area of the inner grain A(t), should shrink according to
\begin{equation}
\frac{d A}{d t}=2\pi M \gamma^{\ast}=2\pi M^{\ast }.
\label{rate_new}
\end{equation}
The reduced mobility therefore corresponds to the shrink rate and, once measured, can be used to determine the MC mobility. To accomplish this, MC simulations were carried out over a wide temperature range $(0<\alpha<2.0)$ with the first nearest neighbors considered. Consistent with results for other lattices ~\cite{Lobkovsky2004285}, the reduced mobility was found to be isotropic at all temperatures. This is shown in Fig. \ref{mc_result} for two representative temperatures. 
The results of the analysis are plotted in Fig. \ref{2d_shrink_rate}, which exhibits a nearly bi-linear dependence of reduced mobility on temperature:
\begin{equation}
2\pi M^{\ast }(\alpha)=\bigg\{
\begin{array}{c}
3.945,\text{ \ \ \ \ \ \ \ \ \ \ \ \ \ \ } \alpha<0.635 \\
5.03-1.74\alpha,\text{\ \ \ \ \ \  } \alpha>0.635.%
\end{array}
\end{equation}
This is a refinement on a result we previously obtained~\cite{PhysRevE.66.061603}.
The MC mobility therefore is:
\begin{equation}
M_{m}(\alpha,\varphi )=\Bigg\{%
\begin{array}{c}
\frac{3.945 J}{2\pi \Delta \sigma^{\ast}},\text{ \ \ \ \ \ \ \ \ \ \ \ }%
\alpha<0.635 \text{\ }\\
\frac{(5.03-1.74\alpha) J}{2\pi \Delta \sigma^{\ast}},\text{ \ \ \ 
}\alpha>0.635%
\end{array}
\label{mobility}
\end{equation}%
with the grain boundary stiffness, $\sigma^{\ast}, $ is given in Eq.~(\ref{stiffness}).

\begin{figure}[ptb]\begin{center}
\includegraphics[width=0.4\textwidth]{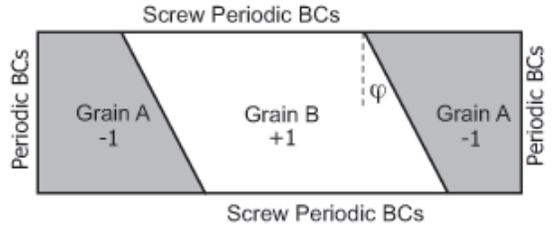}
\caption{
Illustration of the geometry used to measure the mobility of flat grain boundaries driven by an external field. The grid size is $400\times 100$.
}
\label{2d_geometry_new}
\end{center}\end{figure}

\begin{figure}[ptb]\begin{center}
\includegraphics[width=0.4\textwidth]{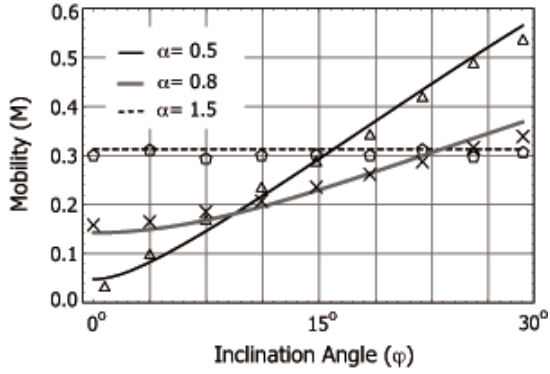}
\caption{
Comparison of grain boundary mobilities measured for two types of driving forces.  The discrete points were obtained from computational experiments with planar boundaries while the solid curves were generated using the capillary result of Eq.~(\ref{mobility}).}
\label{2d_mb_result}
\end{center}\end{figure}

Although this mobility was obtained for a particular type of driving force, we postulate that it should apply to arbitrary driving forces as well.  A second bi-crystal setting was therefore considered in which a planar grain boundary was driven by an external field (Fig. \ref{2d_geometry_new}). 
Two parallel, flat interfaces were constructed with an inclination angle of $\varphi$. Periodic boundary conditions (PBCs) were enforced at left and right, and screw periodic boundary conditions (SPBC) were enforced at top and bottom. A horizontally oriented external field was applied with a magnitude $b_{m}/\Delta^{2}$. The bulk energy density was taken to be the product of this field and the local spin value so that jump in bulk energy across either grain boundary has a magnitude of twice this quantity. A positive value of $b_{m}$ therefore causes the two flat interfaces to move towards each other since the spin value of grain A is -1. The grain boundary velocity can be easily determined by tracking the evolution of grain area, and Eq.~(\ref{bulkeng}) was subsequently used to calculate the grain boundary mobility. Fig.~\ref{2d_mb_result} shows the external field driven mobility with discrete points, and the capillarity driven mobility is shown by solid lines.  Furthermore, the convergence of measured mobility with the two mechanisms confirms the mobility is independent of the type of driving forces. As with the surface energy and grain boundary stiffness, the mobility becomes isotropic as temperature is increased.

\section{Numerical Uncertainty in the Measured Mobility}
\begin{figure}[ptb]\begin{center}
\includegraphics[width=0.4\textwidth]{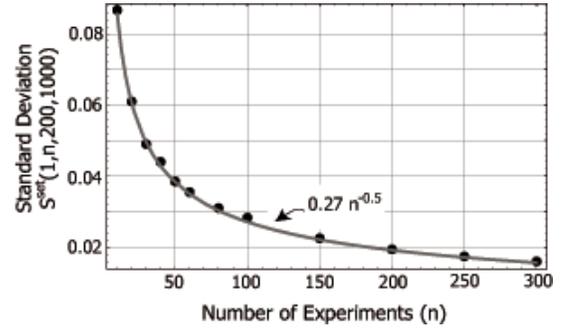}
\caption{
The affect of the number of experiments on the standard deviation of the averaged reduced mobility.
}
\label{numsample_std}
\end{center}\end{figure}

The probabilistic nature of MC simulation calls for a careful analysis of numerical convergence associated with computationally derived relationships such as the reduced mobility. To this end, the capillary shrinking of a circular internal grain was considered once again to determine the influence of number of experiments, $n$, grid size, $L$ and reduced temperature, $\alpha$, on the reduced mobility.  For the purposes of this study, the averaged reduced mobility is taken to be a function of these parameters, $M^{\ast}(\alpha,n,L)$. To generate the fit of  Eq.~(\ref{mobility}), the reduced mobility was estimated using $n=200$ and $L=200$. The reduced mobility was estimated to be
\begin{equation}
M^{\ast}(\alpha,n,L)=\frac{\displaystyle\sum_{i=1}^{n}M^{\ast}_{i}(\alpha,L)}{n},
\label{rmcp}
\end{equation}
where $M^{\ast}_{i}$ stands for the measured reduced mobility associated with the $i^{th}$ experiment. The error bars shown in Fig. \ref{2d_shrink_rate} are the standard deviations $(S)$ of the measurements at each temperature:
\begin{equation}
S(\alpha,n,L)=\sqrt{\frac{\displaystyle\sum_{i=1}^{n}(M^{\ast}_{i}(\alpha,L)-M^{\ast}(\alpha,n,L))^{2}}{n}}.
\label{std}
\end{equation}
This quantifies the statistical character of MC simulation. It is then natural to ask if $200$ samples are enough to generate a mobility function. The issue can be addressed as follows. First, choose the number of experiments, $n$, in a data set and compute the averaged reduced mobility based on this set of data using Eq.(\ref{rmcp}). Then, repeat this process a large number of sets, $m$, and calculate the standard deviation of the averaged reduced mobility using
\begin{equation}
S^{set}(\alpha,n,L,m)=\sqrt{\frac{\displaystyle\sum_{i=1}^{m}(M^{\ast}(i)-M^{\ast, set}(\alpha,n,L,m))^{2}}{m}}.
\end{equation}
Here $M^{\ast}(i)$ is the averaged reduced mobility on the $i^{th}$ set of data and $M^{\ast, set}(\alpha,n,L,m)$ is the mean of averaged reduced mobility of the $m$ data sets: 
\begin{equation}
M^{\ast, set}(\alpha,n,L,m)=\frac{\displaystyle\sum_{i=1}^{m}M^{\ast}(i)}{m}.
\end{equation}
We carried out a set of $1000$ MC simulations at $\alpha=1.0$ with grid size $200\times 200$, and treated the set of reduced mobilities  are viewed as a reservior for statistical analysis. The number of samples were then tested out up to $300$. The discrete points in Fig.~\ref{numsample_std} show the standard deviation of sets of averaged reduced mobility with respect to the number of samples, and the gray line represents the numerical fit provided in the figure.  The set-based standard deviation demonstrates the convergence of the reduced mobility with respect to the number of experiments performed. 

\begin{figure}[ptb]\begin{center}
\includegraphics[width=0.4\textwidth]{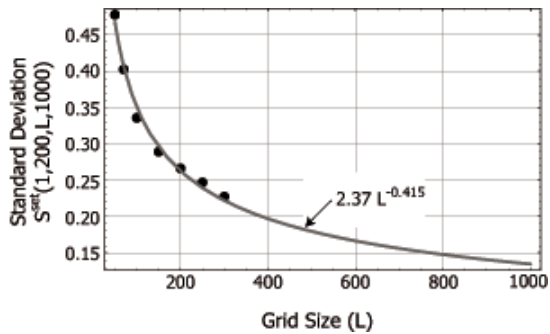}
\caption{
Grid size effect on the standard deviation of independent measured grain boundary reduced mobility.
}
\label{gridsize_std}
\end{center}\end{figure}

The impact of grid size on mobility was taken up next. A set of $200$ experiments were performed with $\alpha=1.0$ for a range of grid sizes. The resulting standard deviations are plotted as discrete points in Fig. \ref{gridsize_std} with a numerical fit shown in gray. This quantifies the degree to which grid size can be used to decrease the variation MC simulation.

\section{Parametric links between MC and SI models and application}
\begin{figure}[ptb]\begin{center}
\includegraphics[width=0.4\textwidth]{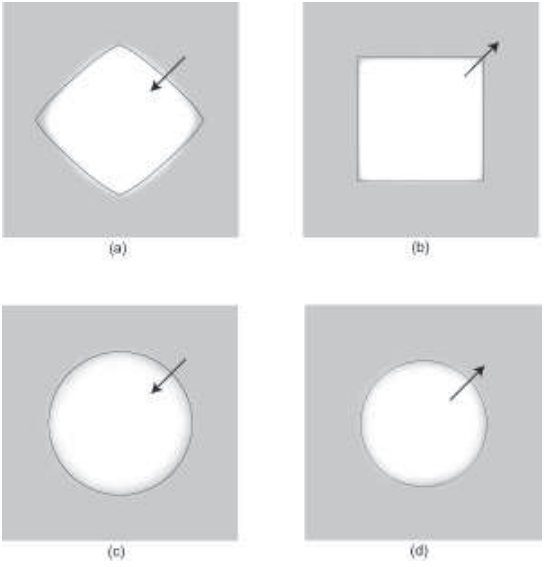}
\caption{
Snapshots of Monte Carlo simulations of inner grain shrinking and growing at (a) $\alpha
=0.5$, and $b_m=0.02$, (b) $\alpha=0.5$, and $b_m= -0.05$, (c) $\alpha=1.5$, and $b_m=0.05$, (d) $\alpha=1.5$,
and $b_m=-0.05$. The grid size is $200\times 200$, $\alpha=0.5$, and 100 CMC simulations were averaged to obtain the results shown.
}
\label{mc_bulk_result}
\end{center}\end{figure}

\begin{figure}[ptb]\begin{center}
\includegraphics[width=0.4\textwidth]{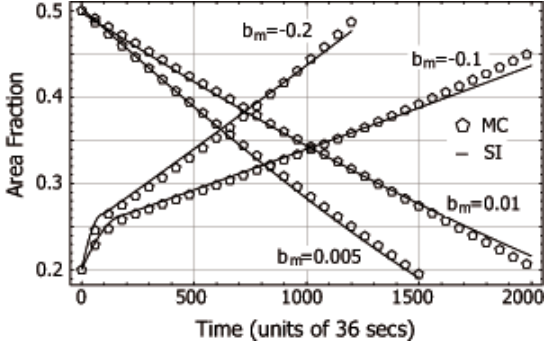}
\caption{
Low temperature CMC predictions are compared with those from the SI model for a range of bulk forces but with the same capillary force. The grid size is $200\times 200$, $\alpha=0.5$, and 100 CMC simulations were averaged to obtain the results shown.
}
\label{2d_gs_lt}
\end{center}\end{figure}

\begin{figure}[ptb]\begin{center}
\includegraphics[width=0.4\textwidth]{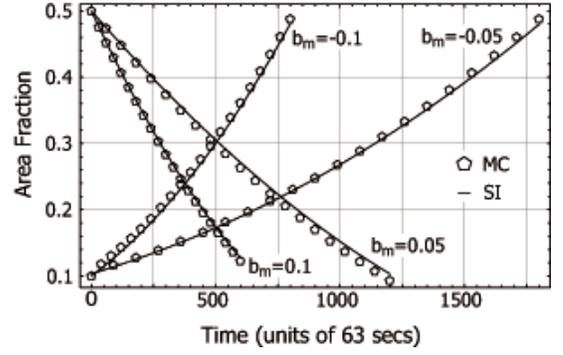}
\caption{
High temperature CMC predictions are compared with those from the SI model for a range of bulk forces but with the same capillary force. The grid size is $200\times 200$, $\alpha=0.5$, and 100 CMC simulations were averaged to obtain the results shown.
}
\label{2d_gs_ht}
\end{center}\end{figure}

The computationally derived MC mobility can be used to establish the parametric links between nondimensional SI and MC models. First recall that the the bulk energy correspondence is
\begin{equation}
b=\frac{2b_{m}}{\Delta ^{2}},  
\label{bulk_link}
\end{equation}
Within a purely capillary setting, the MC kinetic relation can be written as:
\begin{equation}
\frac{d A}{dt}=-(\frac{\Delta M_{m}}{J})(\frac{\Delta ^{2}}{\tau _{mc}}%
)\sigma^{\ast}. 
\label{2dMC_rate}
\end{equation}%
Comparison of Eq.s (\ref{2dMC_rate}) and (\ref{rate_new}) provides the desired link between the nondimensional SI and MC mobilities:
\begin{equation}
M=-\frac{\Delta ^{3}M_{m}}{J\tau _{mc}}.  
\label{2d_link}
\end{equation}
As with the length scales, $\tau_{mc}$ is set to one, which simply means that one MC step is equal to one nondimensional SI time step. Moreover, the grain boundary stiffness is expressed in Eq.~(\ref{stiffness}). The links between nondimensional length, time, energy, and mobility are thus established. To validate these links, MC simulations were performed at a high ($\alpha=1.5$) and a low ($\alpha=0.5$) temperature with a combination of bulk and capillary forces, respectively. Four bulk energies were tested out at each temperature. The MC parameters are then converted to nondimensional SI variables using Eq. (\ref{bulk_link}) and Eq. (\ref{2d_link}). Therefore, the corresponding SI simulations can be carried out. At the high temperature, due to the isotropic grain boundary properties, grain boundary velocity can be written as
\begin{equation}
v=M^{\ast }(\alpha)\kappa +M(\alpha)b,  \label{ht_velocity}
\end{equation}
and the inner grain area changing rate therefore is
\begin{equation}
\frac{\partial A}{\partial t}=2\pi M^{\ast }(\alpha)+2\pi M(\alpha)(\frac{A}{\pi })^{1/2}b.
\end{equation}
In contrast, at the low temperature, anisotropic grain boundary properties must be applied, and the grain boundary velocity now is
\begin{equation}
v=M(\alpha,\varphi )\gamma ^{\ast }(\alpha,\varphi )\kappa +M(\alpha,\varphi )b.
\label{ge_velocity}
\end{equation}
The solution to this differential equation was carried out numerically over one-quarter of the computational domain so that the vertical position of an interface point can be parametrically described by its horizontal position--i.e. $y=f(x)$. The coordinates of points on the grain boundary are described via finite differencing:
\begin{eqnarray}
x_{i}(t+\Delta t) &=&x_{i}(t)+v_{i}\Delta t\cos [\varphi _{i}] \notag\\
y_{i}(t+\Delta t) &=&y_{i}(t)+v_{i}\Delta t\sin [\varphi _{i}]
\end{eqnarray}
These are in terms of the local grain boundary normal speed,
$v _{i}$ is
\begin{equation}
v_{i}=M(T,\varphi _{i})\gamma ^{\ast }(T,\varphi _{i})\kappa
_{i}+M(T,\varphi _{i})b, 
\end{equation}
and the discretized inclination angle:
\begin{equation}
\varphi _{i}=\frac{\pi }{2}+\tan ^{-1}f^{^{\prime }}(x_{i}(t)).
\end{equation}
An expression for the discrete curvature is also required:
\begin{equation}
\kappa _{i}=\frac{f^{^{\prime \prime }}(x_{i}(t))}{(1+f^{^{\prime
}}(x_{i}(t))^{2})^{\frac{3}{2}}}.
\end{equation}
The inner grain area can then be calculated as well:
\begin{eqnarray}
A(t+\Delta t)&=&\sum\limits_{i}\left\vert y_{i}(t+\Delta t)x_{i+1}(t+\Delta t)\right\vert -\notag\\
&&\sum\limits_{i}\left\vert y_{i}(t+\Delta t)x_{i}(t+\Delta t))\right\vert 
\label{diffarea}
\end{eqnarray}
The first set of results are shown in Fig. \ref{mc_bulk_result}, where time slices from four separate experiments are given. The solid curves are the numerical solution to the SI kinetic equation symmetrically reflected from the upper right quadrant into the other three. Agreement between the MC and SI predictions is very good. The top two frames are for low temperature and show how the bulk energy can cause the inner grain to either shrink or grow. These same bulk forces also cause the circular grains to grow anisotropically. The lower two figures are at a much higher temperature and indicate how isotropy is recovered for this regime. Again, the MC model compares well with the SI model.
A more quantitative analysis is provided in Figs. \ref{2d_gs_lt} and \ref{2d_gs_ht}, where the change in inner grain area is plotted versus time for several combinations of bulk and capillary forces. Although the parametric links between SI and MC can be established at an arbitrary MC temperature, high temperature is commonly adopted in MC simulation to reduce the anisotropic lattice effect. These links can also be applied in polycrystalline system.

\section{Calibration of the Monte Carlo Model}
The established parametric links can be used to calibrate the MC algorithm so that MC parameters can be expressed directly in terms of experimental properties. The Calibrated MC (CMC) kinetics then have physical time and length scales. This is achieved using characteristic length, $l_0$, time, $t_0$, and energy $E_0$, to nondimensionalize the physical SI kinetics:
\begin{eqnarray}
b&=&\frac{\overline{b} l_{0}^{2}}{E_{0}} \notag \\
M&=&\frac{\overline{M}E_{0}t_{0}}{l_{0}^{3}}  \notag \\
\sigma^{\ast}&=&\frac{\overline{\sigma^{\ast}}l_{0}}{E_{0}}.
\label{dimnew}
\end{eqnarray}
With the nondimensional correspondence already in place, this allow the MC parameters to be expressed directly in terms of experimental quantities.  We take the physical domain size, $\overline{L}$ to be the characteristic length $(l_{0}=\overline{L})$. Then the product of Eq.~(\ref{dimnew})$_{2}$ and Eq.~(\ref{dimnew})$_{3}$ gives
\begin{equation}
t_{0}=\frac{l_{0}^{2} M^{\ast}(\alpha) \Delta^{2}}{\overline{M^{\ast}}}.
\label{chatime}
\end{equation}
The characteristic energy, $E_{0}$, is derived from the SI grain boundary stiffness, $\overline{\sigma^{\ast}}$: 
\begin{equation}
E_{0}=\frac{\overline{\sigma^{\ast}}\ \l_{0}}{\sigma^{\ast}}.
\label{chaeng}
\end{equation}
As an example, consider a square aluminum sample with a $2$ $mm$ edge length at $500$ $K$. A circular internal grain sits at its center and occupies half of the domain. The interface is taken to have the following isotropic properties~\cite{Humphreys.2004}:
\begin{eqnarray}
\overline{M}&=&10^{-12} \text{ }m^{3}/(J-s),\\
\overline{\sigma^{\ast}}&=&1.0\text{ }J/m.
\end{eqnarray}
The reduced mobility, $\overline{M^{\ast }}$, therefore is
\begin{equation}
\overline{M^{\ast }}=\overline{M} \overline{\sigma}^{\ast}=1.0\times 10^{-12}\text{}m^{2}/s.
\end{equation}
An external field is applied, which causes a bulk energy density difference between the two grains:
\begin{equation}
\overline{b}=1.68\times 10^{4}\text{}J/m^{2}.
\end{equation}%
(This may be viewed as an idealization of dislocation content as well.) The inner grain will shrink due to the combined effect of capillarity and bulk energy difference, and the process can be studied directly using the CMC model. Choose a $200\times 200$ grid so that($\Delta=1/200)$. Also choose an effective MC temperature of $\alpha=1.5$. The high temperature is chosen to correspond to the assumption of isotropy in the experimental data. As mentioned before, the characteristic length is 
\begin{equation}
l_{0}=2.0\times 10^{-3} \text{ }m
\end{equation}
and the characteristic time, $t_{0}$, is
\begin{equation}
t_{0}=\frac{M^{\ast}(1.5) l_{0}^{2}\Delta^{2}}{\overline{M^{\ast}}}= 36 \text{ }s,
\end{equation}
which means one MC step is equal to $36$ seconds in the physical domain. The last value required is the characteristic energy, $E_{0}$:
\begin{equation}
E_{0}=\frac{\overline{\sigma}^{\ast}l_{0}}{\sigma^{\ast}}= 8.4\times 10^{-6} \text{ } J.
\end{equation}
The bulk energy density difference, $\overline{b}$, is nondimensionalized as
\begin{equation}
b=\frac{\overline{b}l_{0}^{2}}{E_{0}}= 8000,
\end{equation}
which is then transformed to a MC bulk energy term
\begin{equation}
b_{m}=\frac{b\Delta^{2}}{2}= 0.1.
\end{equation}
This completes the assignment of MC parameters for this Al example. However, all previous analysis were performed for temperatures sufficiently high that both the experimental and MC kinetics are isotropic. In order to fully evaluate the degree to which the CMC model captures SI behavior, we now take into account the anisotropic properties of grain boundaries; that is, the same physical grain boundary at a relatively low temperature is considered. Now both grain boundary energy and mobility are anisotropic, which can be written as $\overline{M}(\varphi)$ and $\overline{\sigma^{\ast}}(\varphi)$, respectively. Here, the explicit expression of grain boundary mobility and energy is absent due to the lack of experimental data. But the reduced mobility, $\overline{M^{\ast}}$, is still isotropic:
\begin{equation}
\overline{M^{\ast}}=1.0\times 10^{-12}  \text{}m^{2}/s.
\end{equation}
If a low MC temperature $(\alpha=0.5)$ CMC model is applied to mimic this physical process, the characteristic length is still $2\times 10^{-3} m$. The characteristic time now is
\begin{equation}
t_{0}=\frac{M^{\ast}(\alpha) l_{0}^{2}\Delta^{2}}{\overline{M^{\ast}}}= 63 \text{ }s,
\end{equation}
Therefore, one time step in Fig. {\ref{2d_gs_lt} is equal to 63 seconds.

\section{Conclusions}
A two-dimensional, square lattice, Ising model with Metropolis kinetics was used to study grain boundary motion in response to bulk and capillary driving forces. It was found that boundary mobility is not a function of the type of driving force. This is consistent with experimental measurements as well as analytical considerations of triangular lattice systems at low temperatures. The results lend credence to the idea that the independence of mobility and driving force is a universal characteristic of lattice systems.

The identification of a single mobility function allowed us to address a common complaint directed at Monte Carlo investigations: they lack a physical time scale. Simple geometric settings can be used to calibrate the Monte Carlo mobility using experimental data; this endows the Monte Carlo paradigm with a physical time scale. The introduction of more easily obtained length and energy scales admits the representation Monte Carlo parameters directly in terms of experimentally measurable parameters. This new methodology was extensively tested and agreed well with the kinetic results predicted by sharp-interface theory. We refer to such physical probabilistic paradigms as \emph{Calibrated Monte Carlo} models. 

In the present work, the focus was on a single grain boundary within a bi-crystal. However, the resulting Calibrated Monte Carlo model can be immediately applied to polycrystalline material kinetics. The mobility function, and hence the time scale, is the same and only misorientation dependent grain boundary energies is required. The mobility investigation and calibration procedure was restricted to a two-dimensional setting where an analytical expression for the grain boundary free energy made the analysis particularly straightforward. This sets the stage for a consideration of three-dimensional systems, though, where no analytical solution for grain boundary energy is available and where the finite roughening temperature make it especially complicated to work with grain boundary stiffness~\cite{Burkner.1983,PhysRevB.39.7089}.

\section{Acknowledgments}

The research is supported by Sandia National Laboratories. Sandia is a multiprogram laboratory operated by the Sandia Corporation, a Lockheed Martin Company, for the United States Department of Energy under contract DE-AC04-94AL85000. We also acknowledge the Golden Energy Computing Organization at the Colorado School of Mines for the use of resources acquired with financial assistance from the National Science Foundation and the National Renewable Energy Laboratory. 

\section{references}
\bibliographystyle{unsrt}
%\bibliography{mc2d}

\end{document}